\documentclass[11pt]{article}
\usepackage{shonan,natbib}
\usepackage{graphicx}
\usepackage{graphics,float}
\usepackage{epsf}

\usepackage{amsmath}                                
\usepackage{amssymb}
\usepackage{picinpar}                               
\usepackage{relsize,setspace,subfigure}
\usepackage{tabularx}
\usepackage[innercaption]{sidecap}                  
\usepackage[figuresright]{rotating}

\def\d3k{{\displaystyle {\rm d}{\bf k} \over \displaystyle (2\pi)^3}}

\begin{document}
\vspace*{13mm} \noindent \hspace*{13mm}
\begin{minipage}[t]{14cm}
{\large CLUSTERS AND THE COSMIC WEB:\\
Identifying the location of filaments}
\\[13mm]
Rien van de Weygaert\\
{\it Kapteyn Astronomical Institute, University of Groningen, P.O. Box 800, 
  9700 AV Groningen, the Netherlands}\\
\end{minipage}

\section*{Introduction}
\begin{figure*}
\begin{center}
  \vskip -3.5cm
\mbox{\hskip 1.0truecm\includegraphics[width=14.85cm]{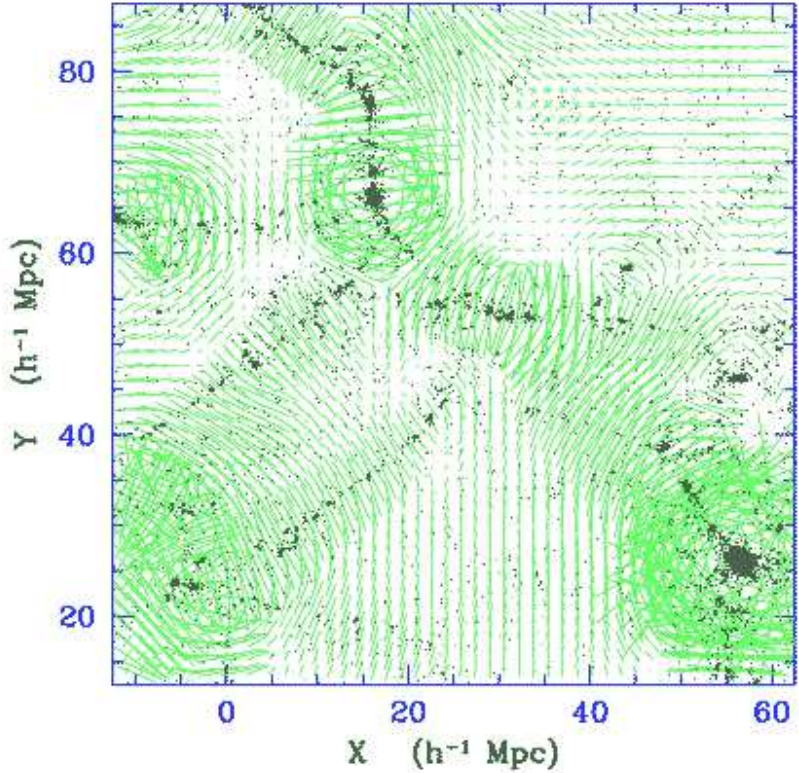}}
  \vskip -1.5cm
\caption{The relation between the {\it cosmic web}, the clusters at the nodes in this network and 
the corresponding compressional tidal field pattern. It shows the matter distribution at the 
present cosmic epoch, along with the (compressional component) tidal field bars in a slice through 
a simulation box containing a realization of cosmic structure formed in an open, $\Omega_{\circ}=0.3$, Universe 
for a CDM structure formation scenario (scale: $R_G=2h^{-1}\hbox{\rm Mpc}$). The frame shows structure in a 
$5h^{-1}\hbox{\rm Mpc}$ thin central slice, on which the related tidal bar configuration is superimposed. 
The matter distribution, displaying a pronounced weblike geometry, is clearly intimately linked with a 
characteristic coherent compressional tidal bar pattern. From:~\cite{ref:weyfoam2002}} 
\label{fig:cosmicwebtide}
\end{center}
\end{figure*}

\noindent The interior of the Universe is permeated by a tenuous space-filling frothy 
network. Welded into a distinctive foamy pattern, galaxies accumulate in walls, filaments 
and dense compact clusters surrounding large near-empty void regions.
As borne out by a large sequence of computer experiments, such weblike patterns in the 
overall cosmic matter distribution do represent a universal but possibly transient phase 
in the gravitationally propelled emergence and evolution of cosmic structure. 
In this contribution we discuss the intimate relationship between the filamentary features 
and the rare dense compact cluster nodes in this network, via the large scale tidal field 
going along with them (see fig~\ref{fig:cosmicwebtide}), following the {\it cosmic web} theory developed 
by~\cite{ref:bondweb1996}. The Megaparsec scale tidal shear pattern is responsible for the 
contraction of matter into filaments, and its link with the cluster locations can be 
understood through the implied quadrupolar mass distribution in which the clusters 
are to be found at the sites of the overdense patches~\citep{ref:weyedbert1996}). The pattern of 
the cosmic web can therefore be largely tied in with the protocluster peaks in the 
primordial density field, and the subsequent nonlinear evolution leads to the aggregation 
of matter into the sharp filamentary network defined by the primordial tidal shear field. 

We present a new technique for tracing the cosmic web, identifying planar walls, elongated 
filaments and cluster nodes in the galaxy distribution. These will allow the practical 
exploitation of the concept of the cosmic web towards identifying and tracing the locations 
of the gaseous WHIM. These methods, the Delaunay Tessellation Field Estimator and the Morphology 
Multiscale Filter~\citep{ref:schaapwey2000,ref:aragon2006}, find their basis in computational 
geometry and visualization. 

\section{Anisotropic Collapse}
\noindent A major characteristic of the formation of cosmic structure in gravitational instability 
scenarios is the tendency of matter concentrations to collapse in an {\it anisotropic} manner. 
In a generic random density field the gravitational force field at any location will be anisotropic. 
For a particular structure the {\it internal} force field of the structure hangs together with the flattening 
of the feature itself. It induces an anisotropic collapse along the main axes of the structure. In reality, 
the internal evolution of the system will be dominated by internal substructure involving a substantial 
measure of orbit crossing. The more quiescent {\it external `background' force field}, the integrated gravitational 
impact of all external density features in the Universe will also be {\it anisotropic}. 

\begin{figure*}
\begin{center}
  \vskip -5.0cm
  \includegraphics[width=14.85cm]{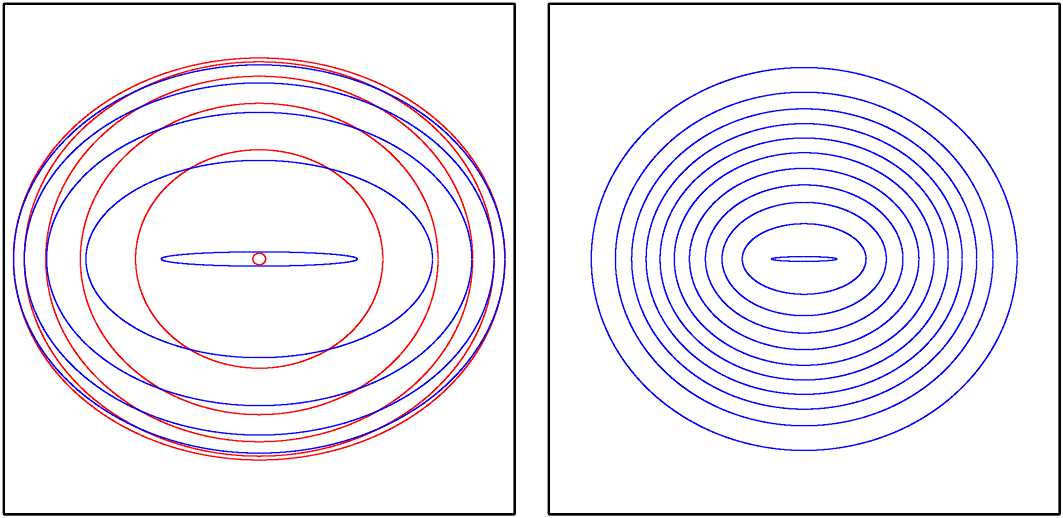}
  \vskip -3.0cm
\caption{The evolution of an overdense homogeneous ellipsoid, with initial axis 
ratio $a_1:a_2:a_3=1.0:0.9:0.9$, embedded in an Einstein-de-Sitter background Universe. 
The two frames show a time sequel of the ellipsoidal configurations attained by the object, 
starting from a near-spherical shape, initially trailing the global cosmic expansion, 
and after reaching a maximum expansion turning around and proceeding inexorably towards 
ultimate collapse as a highly elongated ellipsoid. Left: the evolution depicted in 
physical coordinates. Red contours represent the stages of expansion, blue those of 
the subsequent collapse after turn-around. Right: the evolution of the same object in 
comoving coordinates, a monologous procession through ever more compact and more 
elongated configurations.} 
\label{fig:ellipspeak}
\end{center}
\end{figure*}

In all, the resulting evolution can be most clearly understood in and around a density maximum (or minimum) $\delta$, 
to first order corresponding to the collapse of a homogeneous ellipsoid~\citep{ref:icke1973,ref:eisenstloeb1995, 
ref:peakpatchI}, as illustrated in fig.~\ref{fig:ellipspeak}. The overall characteristics can also be understood 
for the more generic circumstances of a density fluctuation field, where the early phases of the collapse of a 
feature may be approximated by the Zel'dovich deformation tensor $\psi_{mn}$~\citep{ref:zeld1970}. Related to the 
total {\it tidal force field} $T_{mn}$ acting over a patch of density excess $\delta$, including the contributions from 
the local (``internal'') flattening of the density field as well as those generated by external density perturbations, 
the eigenvalues $\lambda_1$, $\lambda_2$ and $\lambda_3$ of the deformation tensor $\psi_{mn}$, 
\begin{eqnarray}
\psi_{mn}\ = \ \frac{\displaystyle {2}}{\displaystyle 3 a^3 \Omega H^2}\,\,
\frac{\displaystyle \partial^2 {\underline \phi}}{\displaystyle \partial q_m 
\partial q_n}\ =\ \frac{\displaystyle 1}{\displaystyle {\scriptstyle{\frac{3}{2}}}\Omega H^2 a} 
\Bigl(\ {\underline T_{mn}}\ +\ {\displaystyle {\frac{1}{2}}}\Omega H^2\ {\underline \delta}\ \delta_{mn}\ \Bigr)\,
\end{eqnarray}
\noindent (underlined quantities are the linearly extrapolated values). Dependent on whether one or more of 
the eigenvalues $\lambda_i>0$, the feature will collapse along one or more directions. The collapse will proceed 
along a sequence of three stages. First, collapse along the direction of the strongest deformation $\lambda_1$. 
The feature will be like a {\it wall}, flattened. If also the second eigenvalue is positive, the object will contract along 
the second direction and an elongated {\it filamentary} structure results. Total collapse will occur if 
also $\lambda_3>0$. 
In $N$-body simulations as well as in galaxy redshift distributions it are in particular the filaments which 
stand out as the most prominent feature of the {\it Cosmic Web}. It even remains unclear whether {\it walls} are 
even present at all. Some argue that once nonlinear clustering sets in the stage in which walls form is of a very 
short duration or does not occur at all: true collapse would proceed along filamentary structures  
\citep{ref:sathya1996, ref:jainedbert1994, ref:huiedbert1996}. Indeed, it can be argued 
that the typical density contours of overdense regions subject to tidal shear constraints are already more 
filamentary than sheet-like in the linear density field, and becomes even more so in the quasi-linear regime 
~\citep{ref:bondweb1996}. In addition, there is also a practical problem in identifying them, due to walls having a 
far lower surface density than the filaments. This is exacerbated as there are hardly any objective feature 
detection techniques available available. Very recent results based upon the 
analysis of an $N$-body simulation of cosmic structure formation by means of the new Multiscale Morphology 
Filter technique indeed identified walls in abundance whether they had not been seen before~\citep{ref:aragon2006}.
 Another indication is that the dissipative gaseous matter within the cosmic web indeed partially aggregates 
in walls with low overdensities~\citep{ref:kangryu2005,ref:kang2006}, arguing for the presence 
of moderate potential wells tied in with dark matter walls. 

\section{Cosmic Web: Tides and Quadrupoles}
\begin{figure*}
\begin{center}
  \vskip -2.0truecm
\mbox{\hskip -0.5truecm\includegraphics[width=16.85cm]{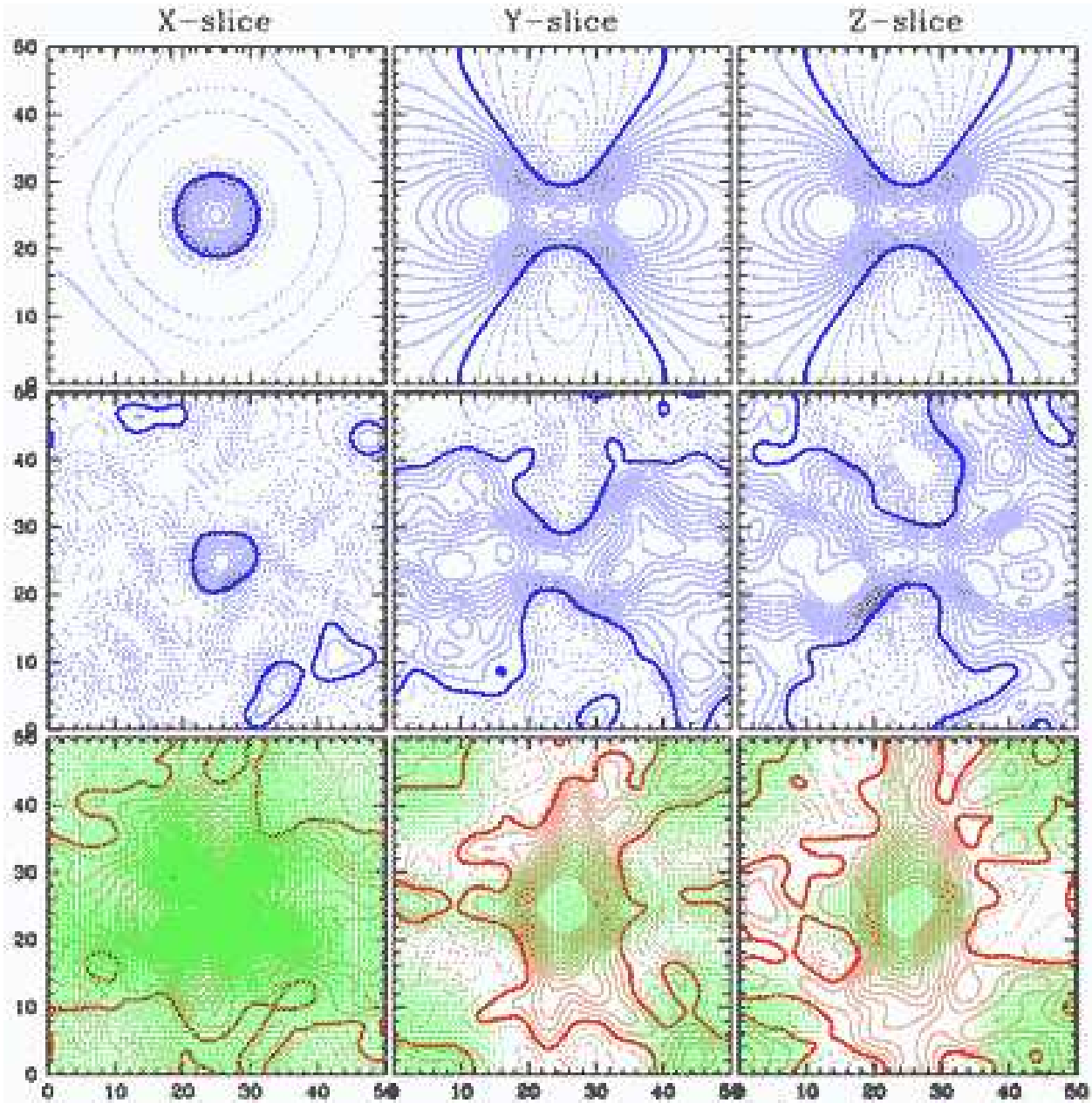}}
\end{center}
\vskip -0.5truecm
\caption{{\small Constrained field construction of initial quadrupolar density pattern in a SCDM 
cosmological scenario. The tidal shear constraint is 
specified at the box centre location, issued on a Gaussian scale of $R_G=2h^{-1}\hbox{\rm Mpc}$ 
and includes a stretching tidal component along the $x$- and $y$-axis acting on a small density 
peak at the centre. Its ramifications are illustrated by means of three mutually perpendicular 
slices through the centre. Top row: the ``mean'' field density pattern, the pure signal 
implied by the specified constraint. Notice the clear quadrupolar pattern in the $y$- and 
$z$-slice,directed along the $x$- and $y$-axis, and the corresponding compact circular density 
contours in the $x$-slice: the precursor of a filament. Central row: the full constrained field realization, 
including a realization of appropriately added SCDM density perturbations. Bottom row: the 
corresponding tidal field pattern in the same three slices. The (red) contours depict the run 
of the tidal field strenght $|T|$, while the (green) tidal bars represent direction and magnitude 
of the {\it compressional} tidal component in each slice (scale: $R_G=2h^{-1}\hbox{\rm Mpc}$). 
From:~\cite{ref:weyfoam2002}}} 
\label{fig:tidecrf}
\vskip -0.2truecm
\end{figure*}

\noindent Bond, Kofman \& Pogosyan (1996) coined the word `cosmic web' in their study 
of the physical content of the web, in which they drew attention to their observation 
that knowledge of the value of the tidal field at a few well-chosen cosmic locations in 
some region would determine the overall outline of the weblike pattern in that region. 
This relation may be traced back to a simple configuration, that of a ``global'' quadrupolar 
matter distribution and the resulting ``local'' tidal shear at a particular location ${\bf r}$. 
Such a quadrupolar primordial matter distribution will almost by default evolve into the canonical 
cluster-filament-cluster configuration which appears so prominently in the cosmic foam 
(see fig~\ref{fig:tidecrf}). For a cosmological (random) matter distribution this close connection 
between local force field and global matter distribution may be elucidated through the expression 
of the tidal tensor in terms of the generating cosmic matter density fluctuation distribution 
$\delta({\bf r})$ \citep{ref:weyedbert1996}:
\begin{eqnarray}
T_{ij}({\bf r})\ = \ {\displaystyle 3 \Omega H^2 \over \displaystyle 8\pi}\,
\int {\rm d}{\bf r}'\,\delta({\bf r}')\ \left\{{\displaystyle 3 (r_i'-r_i)(r_j'-r_j)-
|{\bf r}'-{\bf r}|^2\ \delta_{ij} \over \displaystyle |{\bf r}'-{\bf r}|^5}\right\}\ - \ \nonumber\\
\nonumber\\
\ - \ {\frac{1}{2}}\Omega H^2\ \delta({\bf r},t)\ \delta_{ij}\ \ \ \ \ \ \ \ \ \ \ \ \ \ \ \ \ \ \ \ \ \ \ \ \  
\end{eqnarray}
\section{Constrained Random Field Formalism}
\noindent The set of density field realizations $\delta({\bf r})$ within a sample volume ${\cal V}_s$ 
that would generate a tidal field $T_{ij}$ at location ${\bf r}$ can be inferred from the theory of constrained 
random fields~\citep{ref:edbert1987}. Bertschinger described how a set $\Gamma$ of functional 
field constraints $C_i[f]=c_i,\,(i=1,\ldots,M)$ of a Gaussian random field $f({\bf r},t)$ would translate into  
field configurations for which these constraints would indeed have the specified values $c_i$. Any such 
{\it constrained field realization} $f_c$ can be written as the sum of a {\it mean field} 
${\bar f}({\bf x})=\langle f({\bf x})|\Gamma \rangle$, the ensemble average of all field realizations obeying the 
constraints, and a {\it residual field} $F({\bf x})$, embodying the field fluctuations characterized and 
specified by the power spectrum $P(k)$ of the particular cosmological scenario at hand,  
\begin{equation}
f_c({\bf x})\,=\,{\bar f}({\bf x})\,+\,F({\bf x})\ \\
\end{equation}
\noindent Bertschinger(1987) showed the specific dependence of the mean field on the {\it nature} 
$C_i[f]$ of the constraints as well as their {\it values} $c_i$. In essence the mean field can be seen as the 
weighted sum of the field-constraint correlation functions $\xi_i({\bf x})\equiv\langle f\,C_i\rangle$ (where 
we follow the notation of \cite{ref:hofrib1991}). Each field-constraint correlation function encapsulates the repercussion 
of a specific constraint $C_i[f]$ for a field $f({\bf x})$ throughout the sample volume ${\cal V}_s$. Not surprisingly, 
the field-constraint correlation function for the tidal constraint $T_{ij}$ is a quadrupolar configuration. The weights 
for each of the relevant $\xi_i({\bf x})$ are determined by the value of the constraints, $c_m$, and their mutual 
cross-correlation $\xi_{mn}\equiv \langle C_m C_n\rangle$, 
\begin{eqnarray}
{\bar f}({\bf x})\,=\,\xi_i({\bf x})\,\xi_{ij}^{-1}\,c_j\,.
\end{eqnarray}
\noindent Generating the residual field $F$ is a nontrivial exercise: the specified constraints translate 
into locally fixed phase correlations. This renders a straightforward random phase Gaussian field generation 
procedure unfeasible. Hoffman \& Ribak (1991) pointed out that for a Gaussian random field the sampling is 
straightforward and direct, which greatly faciliated the application of CRFs to cosmological circumstances. 
This greatly facilitated the application of CRFs to complex cosmological issues~\citep{ref:klyphof2003, 
ref:mathis2002}. Van de Weygaert \& Bertschinger (1996), following the Hoffman-Ribak formalism, 
worked out the specific CRF application for the circumstance of sets of local density peak (shape, 
orientation, profile) and gravity field constraints. With most calculations set in Fourier space, the 
constrained field realization for a 
linear cosmological density field with power spectrum $P(k)$ can be computed from the Fourier integral
\begin{eqnarray}
f({\bf x})\,=\,\int\,\d3k\,\biggl[{\hat{\tilde f}}({\bf k}) +
P(k)\,{\hat H}_i({\bf k})\,\xi_{ij}^{-1}\,(c_j-{\tilde c}_j)\biggr]
\,{\rm e}^{-{\rm i}{\bf k}\cdot {\bf x}}
\end{eqnarray}
\noindent with ${\hat H}_i({\bf k})$ the constraint $i$'s kernel (the Fourier transform of 
constraint $C_i[f]$), $c_j$ the value of this constraint, while the tilde indicates it 
concerns a regular unconstrained field realization $\tilde f$.
\begin{figure*}
\vskip -3.0truecm
\begin{center}
\mbox{\hskip -1.5truecm\includegraphics[width=19.85cm]{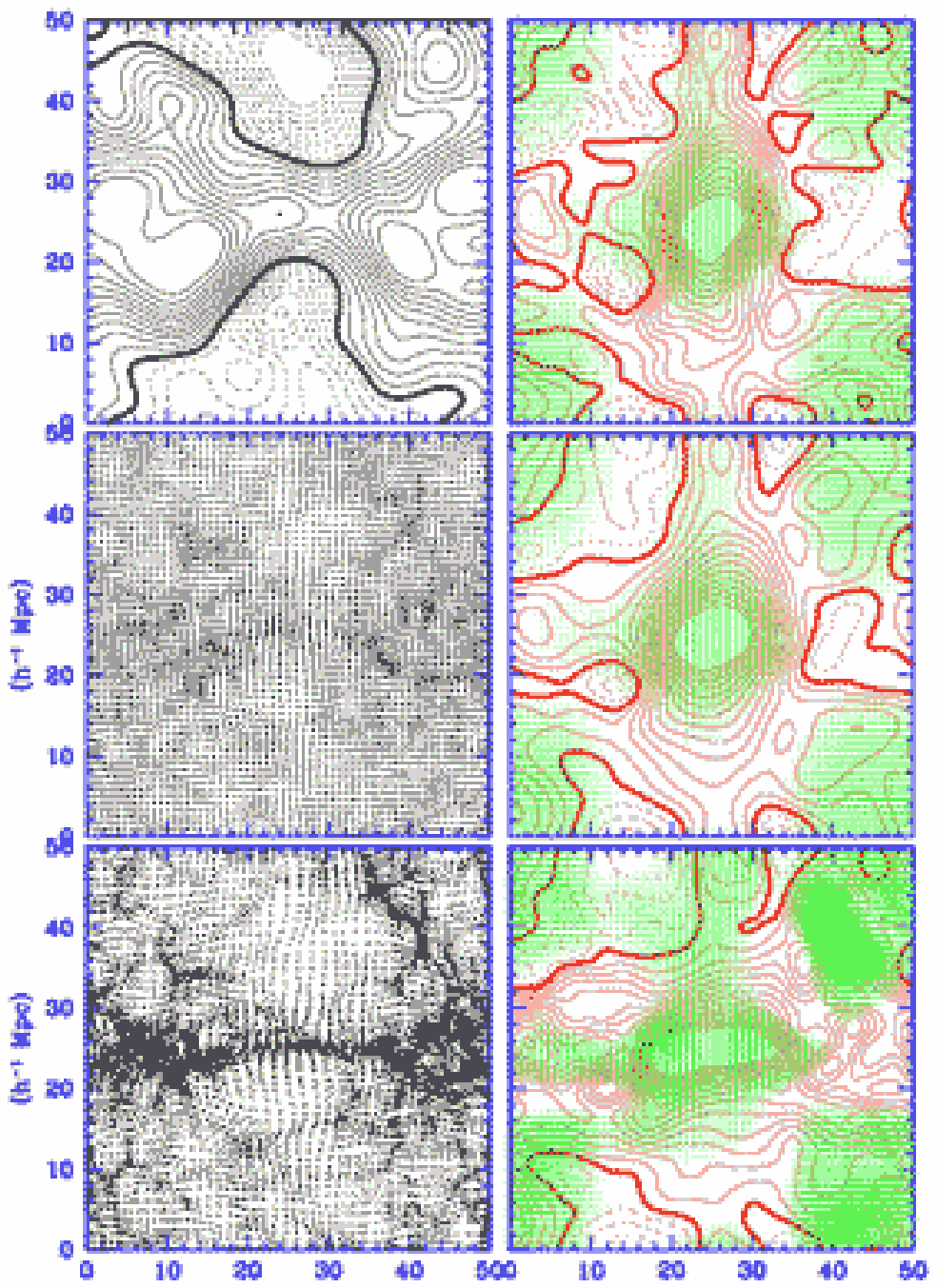}}
\caption{The emergence of a filament in an SCDM structure formation scenario. Lefthand column: 
density/particle distribution in $z$-slice through the centre of the simulation box. Righthand 
column: the corresponding tidal field configurations, represented through the full tidal 
field strength $|T|$ contour maps (red), as well as the corresponding compressional tidal 
bars (scale: $R_G=2h^{-1}\hbox{\rm Mpc}$). From top to bottom: primordial field, 
$a=0.2$ (visible emergence filament), present epoch. Note the formation of the filament at the 
site where the tidal forces peaked in strength, with a tidal pattern whose topology remains 
roughly similar. From:~\cite{ref:weyfoam2002}} 
\label{fig:filcrf}
\end{center}
\end{figure*}
While the CRF formalism is rather straightforward for idealized linear constraints, reality is less 
forthcoming. If the constraints are based on measured data these will in general be 
noisy, sparse and incomplete. Wiener filtering will be able to deal with such a situation and 
reconstruct the implied {\it mean field}, at the cost of losing signal proportional to the loss
in data quality (see e.g. \cite{zaroubi1995}). A major practical limitation concerns the 
condition that the constrained field is Gaussian. For more generic nonlinear clustering situations 
the formalism is in need of additional modifications. For specific situations this may be feasible 
~\citep{ref:sheth1995}, but for more generic circumstances this is less obvious (however, see
~\cite{ref:joneswey2006}).

\section{Tidal connections: Filaments and Clusters}
\noindent One of the major virtues of the {\it constrained random field} construction technique is that it offers the 
instrument for translating locally specified quantities into the corresponding implied global matter 
distributions for a given structure formation scenario. In principle, the choice of possible implied matter 
distribution configurations is limitless, yet it gets substantially curtailed by the statistical nature of 
its density fluctuations, the coherence scale of the matter distribution and hence of the generated force field 
as well as the noise characteristics over the various spatial scales, both set by the power 
spectrum of fluctuations. Van de Weygaert \& Bertschinger (1996) illustrate the repercussion of a specified constraint 
on the value of the tidal shear at some specific location. Figure~\ref{fig:filcrf} shows the result of such a tidal 
constraint. It provides a 3-D impression of the structure in the region immediately surrounding the location of the 
specified shear. We have crudely included the concept of ``external'' by (spherically) filtering the field on a 
(rather arbitrary) scale of $2h^{-1}\hbox{\rm Mpc}$. 

The {\it mean field} ${\bar f}$ of the specified constraints (top panels) represents a clear depiction of the average 
density field configuration inducing the specified tidal tensor: the constraint works out into a perfect global quadrupolar 
field. Superimposing the {\it residual} power spectrum fluctuations $F$, whose amplitude is modified by the local correlation 
with the specified constraints, results into a representative individual realization of a matter density distribution that 
would induce the specified constraint (second row). The close affiliation with a strong anisotropic force field, within 
the surrounding region, can be directly observed from the lower row of corresponding slices. The contour maps, 
indicating the total tidal field strength, reveal that the constraints correspond to a tidal field elongated along 
the axis of the box with a maximum tidal strength at the centre of the box. Along the full length of the filament we 
observe a coherent pattern of strong compressional forces perpendicular to its axis\footnote{on the basis of the effect 
of a tidal field, we may distinguish at any one location between ``compressional'' and ``dilational'' 
components. Along the direction of a ``compressional'' tidal component $T_{c}$ (for which  
$T_c<0.0$) the resulting force field will lead to contraction, pulling together the matter 
currents. The ``dilational'' (or ``stretching'') tidal component $T_d$, on the other hand, 
represents the direction along which matter currents tend to get stretched as $T_d>0$. Note that within a plane, 
cutting through the 3-D tidal ``ellipsoid'', the tidal field can consist of two compressional 
components, two dilational ones or -- the most frequently encountered situation -- of one 
dilational and one compressional component.}. 
\bigskip
Assessing the evolution of the spatial matter distribution in and around the (proto)filament, see Fig.~\ref{fig:filcrf},  
demonstrates the intimate correlation between the anisotropy in the cosmic force field and the presence of 
strongly anisotropic features. It shows the emergence of a filament in a CDM structure formation scenario, 
with the density//particle distribution along the ``spine'' of the emerging filament in the lefthand column and 
the corresponding tidal configuration (full tidal field strength contour map as well as corresponding bars of compressional 
tidal component) in the righthand column. The top row corresponds to the primordial cosmic conditions, the centre row to 
$a=0.2$ and the bottom row to $a=0.8$. At $a=0.2$ we can clearly recognize the onset of the emerging filament, 
which at $a=0.8$ has emerged as the dominant feature in the mass distribution. Two striking aspects of the 
depicted evolution are particularly relevant for this contribution:

\vfill\eject
\begin{itemize}
\item{} Two {\it massive clusters} emerge on either side of the filament. These matter assemblies, in 
conjunction with the correspondingly large underdense volumes surrounding the filament perpendicular to its 
spinal axis, define a roughly quadrupolar density field and are a natural consequence of the primordial 
density field suggested by the central tidal force field constraint (fig.~\ref{fig:tidecrf}). 
\item{} The strong correlation between the compressional component of the tidal field and the presence 
of a dense filamentary feature suggests a strong causal link (fig.~\ref{fig:cosmicwebtide},~\ref{fig:filcrf}). 
A comparison between the 
evolving cosmic web and the corresponding tidal force field, specifically of its compressional components, 
does suggest an intimate 
link. While the spatial pattern of the tidal field remains quite close to its primordial configuration, 
we see the formation of the filament precisely there where the primordial compressional field is very 
strong and coherent. In other words, it is as if {\it the primordial mapping of the compressional tidal 
component represents a prediction for the locus of the main cosmic web features}. The gradual emergence 
of one particular filament is seemingly predestinated by the tidal field configuration.
\end{itemize}

\section{Tidal connections: Clusters and the Web}
Inverting the relation between clusters and the cosmic web, we may investigate the repercussions of 
imposing the locations and nature of cluster nodes to trace out the implied cosmic web. This has been 
described in detail in \cite{ref:bondweb1996}. Clusters are defined according to the peak-patch 
formalism of Bond \& Myers (1996): they are peaks in the primordial Gaussian field, identified 
with the peak on the largest smoothing scale $R_G$ on which they have collapsed along all three directions 
(according to the homogeneous ellipsoidal model). 

As argued in the above, the presence of two protocluster peaks may imply that the tidal shear field 
configuration in between the peaks is such that a filament will form along the axis connecting the 
two clusters. The strength of the filamentary bridge depends on the distance between the two peaks. 
Its coherence and strength are set by the field-constraint correlation function $\xi_k({\bf r})=\langle 
\delta({\bf r}) T_{ij}({\bf r_T})\rangle$ between the density field and the tidal 
shear. The strength of the correlation will depend strongly on both orientation and of the clusters and 
their mutual distance (\cite{ref:bondweb1996}, also see \cite{ref:weyedbert1996}). This was indeed confirmed in 
a study of fully evolved intracluster filaments in GIF simulations~\citep{ref:colberg2005}. In the observed galaxy 
distribution, ``superclusters'' are therefore filamentary cluster-cluster bridges, and the most pronounced ones 
will be found between clusters of galaxies that are close together and which are aligned with each other. 
Very pronounced galaxy filaments, the Pisces-Perseus supercluster chain is a telling example, are therefore 
almost inescapably tied in with a high concentration of rich galaxy clusters.

These observations can provide the path towards an efficient tracing of weblike patterns at 
higher redshifts. Clusters of galaxies are observable out to high redshifts $z>1$. Using 
the cluster distribution within a particular cosmic region as input, the CRF technique will allow 
the reconstruction of the corresponding filamentary weblike patterns. This in turn may focus 
attention on the highest density patches for tracing the intergalactic gas and thus suggest 
an efficient observational technique. In turn, it will allow a test of structure formation. 
\begin{figure*}
\begin{center}
\vskip 0.0truecm
\mbox{\hskip 0.0truecm\includegraphics[height=23.0cm]{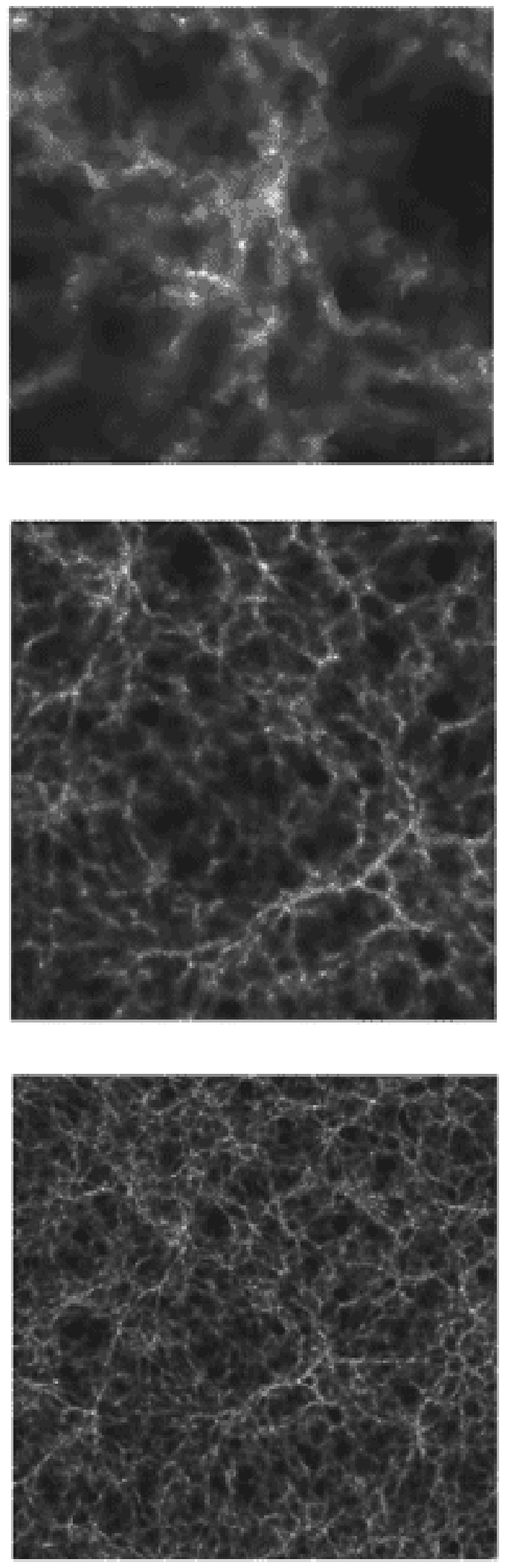}}
\vskip 0.0truecm
\caption{The Cosmic Web in a box: GIF N-body simulation of structure formation in a $\Lambda$CDM 
cosmology. Three consecutive zoom-ins onto a central slice through the simulation box. Courtesy: 
Willem Schaap. }
\label{fig:gifdtfexpanel}
\end{center}
\end{figure*}
\begin{figure}
\begin{center}
\vskip 0.0truecm
\mbox{\hskip 0.0truecm\includegraphics[width=23.0cm,angle=90.0]{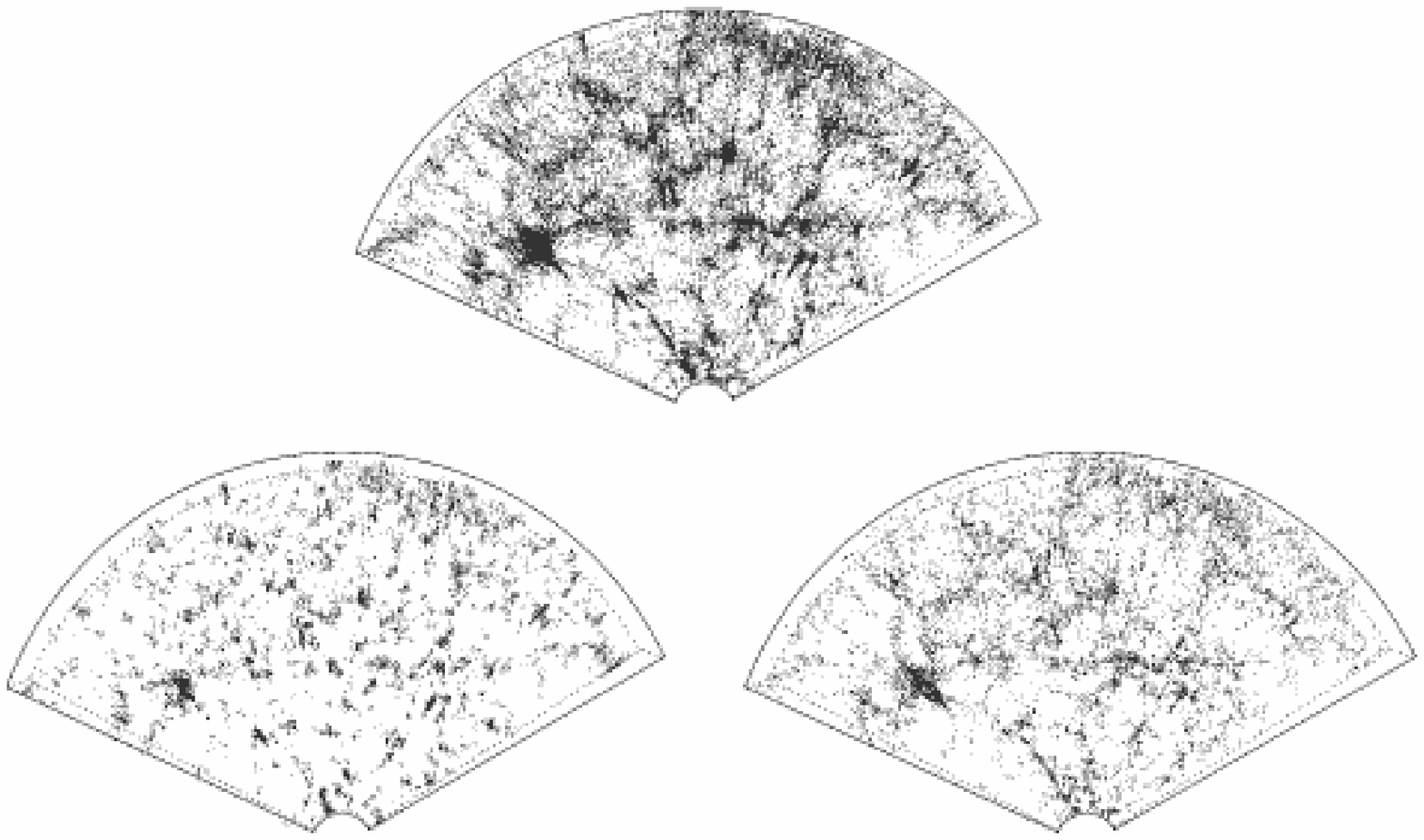}}
\vskip 0.0truecm
\caption{MMF (multiscale morphology filter) analysis of a slice in te SDSS survey. 
From the full galaxy distribution (left), the MMF identifies the galaxies belonging 
to clusters (top right) and to filaments (bottom right). From:~\cite{ref:aragon2006}}
\label{fig:mmfsdss}
\end{center}
\end{figure}
\section{Clusters and Filaments: Identification}
Astronomical applications are usually based upon a set of user-defined filter functions. 
Nearly without exception the definition of these include pre-conceived knowledge about the features 
one is looking for. A telling example is the use of a Gaussian filter. This filter will 
suppress the presence of any structures on a scale smaller than the characteristic filter 
scale. Moreover, nearly always it is a spherically defined filter which tends to smooth out 
any existing anisotropies. Such procedures may be justified in situations in which 
we are particularly interested in objects of that size or in which physical 
understanding suggests the smoothing scale to be of particular significance. 
On the other hand, they may be crucially inept in situations of which we do not know 
in advance the properties of the matter distribution. The gravitational clustering process in 
the case of hierarchical cosmic structure formation scenarios is a particularly notorious 
case. As it includes structures over a vast range of scales and displays a rich palet 
of geometries and patterns any filter design tends to involve a discrimination against 
one or more -- and possibly interesting -- characteristics of the cosmic matter 
distribution it would be preferrable to define filter and reconstruction procedures 
that tend to be defined by the discrete point process itself. 

Here we exploit the potential of spatial {\bf tessellations} as a means of estimating and interpolating 
discrete point samples into continuous field reconstructions, in particular that of {\it Voronoi} 
and {\it Delaunay tessellations} . Both tessellations -- each others {\it dual} -- are fundamental 
concepts in the field of stochastic geometry. They formed the basis of the technique of 
the Delaunay Tessellation Field Estimator (DTFE), defined and introduced by Schaap \& van de Weygaert 
(2000). The DTFE technique is capable of delineating the hierarchical and anisotropic nature of spatial point 
distributions and in outlining the presence and shape of voidlike regions. It is precisely this which marks the 
spatial structure of the cosmic web. DTFE is based upon the use of the Voronoi and Delaunay tessellations of a 
given spatial point distribution to form the basis of a natural, fully self-adaptive filter for 
discretely sampled fields in which the Delaunay tessellations are used as multidimensional 
interpolation intervals. DTFE exploits two particular properties of Voronoi and Delaunay tessellations. 
The tessellations are very sensitive to the local point density, in that the volume of the tessellation cells 
is a strong function of the local (physical) density. The DTFE method uses this fact to define a local estimate 
of the density. It subsequently uses the adaptive and minimum triangulation properties of Delaunay tessellations 
to use them as adaptive spatial interpolation intervals for irregular point distributions. In this it is the 
first order version of the {\it Natural Neighbour method} (NN method). The theoretical basis for the NN method, a 
generic smooth and local higher order spatial interpolation technique developed by experts in the field of 
computational geometry, has been worked out in great detail by \citep{ref:sibson1980,ref:sibson1981,ref:watson1992}. 
As has been demonstrated by telling examples in geophysics~\citep{ref:sambridge1995} and solid mechanics 
and engineering~\citep{ref:sukumar1998} NN methods hold tremendous potential for grid-independent analysis 
and computations. 
The performance of DTFE may be appreciated from fig.~\ref{fig:gifdtfexpanel}. It clearly 
produces a continuous density field that includes, both qualititatively and quantitatively, 
all essential information of the underlying cosmic web. 
\bigskip
Based on this optimized cosmic web reconstruction Arag\'on-Calvo et al.~\cite{ref:aragon2006} developed 
the Multiscale Morphology Filtere technique, particularly oriented towards recognizing and identifying the major 
characteristic elements in the Megaparsec matter density field. The MMF yields a unique framework for 
the combined identification of dense, compact bloblike clusters, of the salient and moderately dense 
elongated filaments and of tenuous planar walls. Of fundamental importance is the use of a morphologically 
unbiased and optimized continuous density field retaining all features visible in a discrete galaxy or 
particle distribution. This is accomplished by means of DTFE.
 
It is based upon an assessment of the coherence of a density (or intensity) field along a range of 
spatial scales and with the virtue of providing a generic framework for characterizing the local 
morphology of the density field and enabling the selection of those morphological features which 
the analysis at hand seeks to study. The technology finds its origin in computer vision research and has been 
optimized within the context of feature detections in medical imaging. \cite{ref:frangi1998} 
and \cite{ref:sato1998} presented its operation for the specific situation of detecting the web of 
blood vessels in a medical image, a notoriously complex pattern of elongated tenuous features whose 
branching make it closely resemble a fractal network. Arag\'on-Calvo et al.~\citep{ref:aragon2006} translated, 
extended and optimized this technology towards the recognition of the major characteristic structural elements 
in the Megaparsec matter distribution of a method finding its origin in computer vision research.  

The first applications to the galaxy distribution in the Sloan Digital Sky Survey produced 
a spectacular result: an objective cluster and filament catalog. Equally encouraging were 
the results of an application to a modelled galaxy distribution in the Millennium Simulation. 

\section{Acknowledgement}

\bigskip The author would like to thank Dick Bond for many years of inspiration 
and encouragement in studying the cosmic web and constrained fields. I am grateful 
to both Bernard Jones and Manolis Plionis for providing substantial feedback on various 
relevant issues. In addition, Willem Schaap and Miguel Arag\'on-Calvo are acknowledged 
for essential contributions to the presented work.

\end{document}